\documentclass[pra,twocolumn,tightenlines,nofootinbib]{revtex4}
\usepackage{bm,dcolumn,amsmath,graphicx}

\begin{document}
\title{Magic wavelengths for optical cooling and trapping of lithium}

\author{M. S. Safronova$^1$}
\author{U. I. Safronova$^{2,3}$}
\author{Charles W. Clark$^{4}$}
\affiliation{}

\affiliation {$^1$Department of Physics and Astronomy, 217 Sharp
Lab, University of Delaware, Newark, Delaware 19716\\
$^2$Physics Department, University of Nevada, Reno, Nevada 89557,
\\$^3$Department of Physics,  University of Notre Dame,
Notre Dame, IN 46556
 \\$^4$Joint Quantum Institute, National Institute of Standards and Technology
and the University of Maryland, Gaithersburg, Maryland 20899-8410, USA}

\date{\today}

\begin{abstract}

Using first-principles calculations, we identify magic wavelengths
$\lambda$ for the $2s-2p$ and $2s-3p$ transitions in lithium. The
$ns$ and $np$ atomic levels have the same ac Stark shifts at the
corresponding magic wavelength, which facilitates state-insensitive
optical cooling and trapping. Tune-out wavelengths for which the
ground-state frequency-dependent polarizability vanishes are also
calculated. Differences of these
 wavelengths between $^6$Li and $^7$Li are
reported. Our approach uses high-precision, relativistic all-order
 methods in which all single, double, and partial triple excitations
of the Dirac-Fock wave functions
 are included to all orders of perturbation theory. Recommended values are provided
 for a large number of Li
electric-dipole matrix elements. Static polarizabilities for the
$2s$, $2p$, $3s$, $3p$, and $3d$ levels are compared with other
theory and experiment where available.  Uncertainties of all
recommended values are estimated.  The magic wavelengths for the uv
$2s-3p$ transition are of particular interest for the production of a
quantum gas of lithium [Duarte \textit{et al.}, Phys. Rev. A
\textbf{84}, 061406R (2011)].
\end{abstract}
\maketitle
\section{Introduction}

The alkali atoms, with one electron outside a closed shell core, have an oscillator strength sum,
 $\Sigma f = 1$, to be
distributed among all optical transitions from the ground state at energies
 below the threshold for core excitation ~\cite{RevModPhys40441}. Most of this oscillator strength is concentrated in the lowest, $ns - np$, resonance transitions, where $n =2$ for Li, 3 for Na, etc. The higher resonance transitions, $ns - n'p$ with $n' > n$, are thus usually much weaker than their conterparts in non-alkali atoms, and their natural linewidths are narrower. In some cases this is accompanied by anomalous
 intensity ratios of the fine-structure doublet lines, a phenomenon first explained by Fermi \cite{1930ZPhy680F}.

Recent experiments in $^6$Li \cite{DuaHarHit11} and $^{40}$K
\cite{MckJerFin11} have shown the advantages of laser cooling with
higher resonance lines for reducing the temperature and increasing
the phase space density of optically trapped alkali atoms. The narrow
linewidths of these transitions reduce the Doppler cooling limit
compared to that of the lowest resonance line.  Two direct tests were
made using similar schemes: laser cooling of the gas in a
magneto-optical trap (MOT) using the usual $ns\rightarrow np_{3/2}$
D2 transition (the wavelengths of which are 671~nm for $^6$Li and
767~nm for $^{40}$K ), followed by transfer of the gas into a MOT
operating at the $ns\rightarrow (n+1)p_{3/2}$ UV line (respectively
323~nm and 405~nm).  In both cases, temperature reductions by a
factor of about five and phase-space density increases by at least a
factor of ten were observed.

 In this paper we discuss aspects of this cooling scheme as applied to lithium.
 In order to be able to continue to laser cool on the UV transition
   when a dipole optical trap is turned off, the ac Stark shifts of the $2s$ and  $3p_{3/2}$
  levels due to trap light have to be nearly the same, resulting
  in a sufficiently small differential ac Stark shift on the
   cooling transition to allow for efficient and uniform cooling ~\cite{DuaHarHit11}.
  The ac Stark shifts of these states are generally different and may
  be of different signs, leading to heating.
The same problem, i.e. different Stark shifts of two states,
    affects optical frequency standards based on atoms
   trapped in optical lattices, because it can introduce a significant
   dependence of the measured frequency of the clock transition upon the lattice wavelength.
  The idea of using a trapping laser tuned to a ``magic'' wavelength, $\lambda_{\rm{magic}}$,
at which the ac Stark shift of the clock transition vanishes, was first proposed in
Refs.~\cite{KatIdoKuw99,YeVerKim99}. The use of the magic
wavelengths is also important
 in order to trap and control neutral atoms inside high-Q cavities in
 the strong coupling regime with minimum decoherence
for quantum computation and communication schemes \cite{MckBucBoo03}.

  The goal of the present work is to provide the list of precise
  magic wavelengths for Li UV $2s-3p_j$ transitions in convenient wavelength regions.
   We also provide a list of
available magic wavelength for the $2s-2p_j$ transitions, calculate dc and ac polarizabilities for several low-lying states,  and provide recommended
values for a number of relevant electric-dipole transitions.
 We present results for both $^6$Li and $^7$Li to illustrate the possibilities
  for differential light shifts between the two isotopes.

The magic wavelengths for a specific transition are located by
calculating the ac polarizabilities of the lower and upper states and
finding their crossing points. Magic wavelengths for lowest $np-ns$
transitions in alkali-metal atoms from Na to Cs have been previously
calculated in \cite{AroSafCla07}, using a relativistic linearized
coupled-cluster method. The data in Ref.~\cite{AroSafCla07} provided
a wide range of magic wavelengths for alkali-metal atoms.  A
bichromatic scheme for state-insensitive optical trapping of Rb atom
was explored in Ref.~\cite{AroSafCla10}. In the case of Rb, the magic
wavelengths associated with monochromatic trapping were sparse and
relatively inconvenient. The bichromatic approach yielded a number of
promising magic wavelength pairs.

We have also carried out calculations of the dc polarizabilities of
the $2s$, $2p_{j}$, $3s$, $3p_{j}$, and $3d_{j}$ states to compare
with available experimental and high-precision theoretical values. We
refer the reader to the recent review \cite{MitSafCla10} for
extensive comparison of various results for Li polarizabilities.
Here, we provide comparison of our results with the most recent
calculations carried out using Hylleraas basis functions
\cite{TanYanShi09,TanYanShi10,TanZhaYan10,PucKedPac11,PucKcePac12}
since this approach is expected to  produce the most accurate
recommended values. Comparison with the recent coupled-cluster
calculation of \cite{WanSahTim10} and configuration interaction
calculations with core polarization (CICP) \cite{TanZhaYan10} is also
included.

We also provide values for the tune-out wavelengths,
$\lambda_{\rm{zero}}$, for which the ground-state frequency-dependent
polarizability of Li atoms vanishes.  At these wavelengths, an atom
experiences no dipole force and thus is unaffected by the presence
 of an optical lattice. These wavelengths were first discussed
  by LeBlanc and Thywissen \cite{LebThy07}, and have recently been calculated
   for alkali-metal atoms from Li to Cs in \cite{AroSafCla11}.
  Our present work provides accurate predictions of the
tune-out wavelengths $\lambda_{\rm{zero}}$ for both $^6$Li and $^7$Li
and explores the differences between the first tune-out wavelengths
for these isotopes in detail.

 We start with the  discussions of the calculation of
electric-dipole
 matric elements, static and dynamic polarizabilities as well as their
uncertainties.  The static values are compared with other theory and
experiment in Section~\ref{sec2}. The magic wavelengths are discussed
in Section~\ref{sec3}.

\section{Method}

The background to our approach to calculation of atomic polarizabilities
was discussed  in Refs.~\cite{AroSafCla07,JonSafDer08,AroSafCla11}.
  Here, we summarize
points salient to the present work. The frequency-dependent scalar
polarizability, $\alpha(\omega)$, of an alkali-metal atom in its
ground state $ v$ may be separated into a contribution from the core
electrons, $\alpha_{\rm{core}}$, a core modification due to the
valence electron, $\alpha_{vc}$, and a contribution from the valence
electron, $\alpha^v(\omega)$.  The core polarizability depends weakly
on $\omega$ for the frequencies treated here, since core electrons
have excitation energies in the far-ultraviolet region of the
spectrum. Therefore, we approximate the core polarizability by its dc
value as calculated in the random-phase approximation (RPA). The
accuracy of the RPA approach has been discussed in
~\cite{MitSafCla10}. The core polarizability is corrected for Pauli
blocking of core-valence
 excitations by introducing an extra term $\alpha_{vc}$.
For consistency, this is also calculated in RPA.
 The valence contribution to frequency-dependent scalar $\alpha_0$ and
 tensor $\alpha_2$ polarizabilities is
evaluated as the sum over intermediate $k$ states allowed by the electric-dipole transition rules~\cite{MitSafCla10}
\begin{eqnarray}
    \alpha_{0}^v(\omega)&=&\frac{2}{3(2j_v+1)}\sum_k\frac{{\left\langle k\left\|d\right\|v\right\rangle}^2(E_k-E_v)}{     (E_k-E_v)^2-\omega^2}, \label{eq-1} \nonumber \\
    \alpha_{2}^v(\omega)&=&-4C\sum_k(-1)^{j_v+j_k+1}
            \left\{
                    \begin{array}{ccc}
                    j_v & 1 & j_k \\
                    1 & j_v & 2 \\
                    \end{array}
            \right\} \nonumber \\
      & &\times \frac{{\left\langle
            k\left\|d\right\|v\right\rangle}^2(E_k-E_v)}{
            (E_k-E_v)^2-\omega^2} \label{eq-pol},
\end{eqnarray}
             where $C$ is given by
\begin{equation}
            C =
                \left(\frac{5j_v(2j_v-1)}{6(j_v+1)(2j_v+1)(2j_v+3)}\right)^{1/2} \nonumber
\end{equation}
and ${\left\langle k\left\|d\right\|v\right\rangle}$ are the reduced electric-dipole matrix elements. In these equations, $\omega$ is assumed to be
at least several linewidths off resonance with the corresponding transitions. Linear polarization is assumed in all calculation.

 Unless stated otherwise, we use the conventional system of atomic units,
a.u., in which $e, m_{\rm e}$, $4\pi \epsilon_0$ and the reduced Planck constant $\hbar$ have the numerical value 1.  Polarizability in a.u. has the
dimension of volume, and its numerical values presented here are expressed in units of $a^3_0$, where $a_0\approx0.052918$~nm is the Bohr radius. The
atomic units for $\alpha$ can be converted to SI units via
 $\alpha/h$~[Hz/(V/m)$^2$]=2.48832$\times10^{-8}\alpha$~[a.u.], where
 the conversion coefficient is $4\pi \epsilon_0 a^3_0/h$ and the
 Planck constant $h$ is factored out.
\begin{table}[ht]
\caption[]{ Absolute values of the reduced electric-dipole matrix elements in
 Li and their uncertainties. Units: a.u.} \label{tab1}
\begin{ruledtabular}
\begin{tabular}{lclc}
   \multicolumn{1}{c}{Transition} & \multicolumn{1}{c}{Value} &
  \multicolumn{1}{c}{Transition} & \multicolumn{1}{c}{Value} \\
\hline
$ 2s    -     2p_{1/2}$&  3.3169(6)  &   $                 $&                  \\
$ 2s    -     3p_{1/2}$&  0.183(3)   &   $  3s    -     3p_{1/2}$&    8.467(2)      \\
$ 2s    -     4p_{1/2}$&  0.160(1)   &   $  3s    -     4p_{1/2}$&    0.0320(6)     \\
$ 2s    -     5p_{1/2}$&  0.1198(9)  &   $  3s    -     5p_{1/2}$&    0.1544(2)     \\
$ 2s    -     6p_{1/2}$&  0.0925(7)  &   $  3s    -     6p_{1/2}$&    0.138(1)      \\
$ 2s    -     7p_{1/2}$&  0.0737(5)  &   $  3s    -     7p_{1/2}$&    0.1136(6)     \\  [0.2pc]
$ 2s    -     2p_{3/2}$&   4.6909(8) &   $                 $&                 \\
$ 2s    -     3p_{3/2}$&   0.259(4)  &   $  3s   -      3p_{3/2}$&    11.975(2)    \\
$ 2s    -     4p_{3/2}$&   0.226(2)  &   $  3s   -      4p_{3/2}$&    0.045(1)     \\
$ 2s    -     5p_{3/2}$&   0.169(1)  &   $  3s   -      5p_{3/2}$&    0.2184(5)    \\
$ 2s    -     6p_{3/2}$&   0.131(1)  &   $  3s   -      6p_{3/2}$&    0.195(2)     \\
$ 2s    -     7p_{3/2}$&   0.1042(7) &   $  3s   -      7p_{3/2}$&    0.1607(9)    \\ [0.2pc]
$ 2p_{1/2}  -  3s$&  2.4326(5)  &   $                 $&              \\
$ 2p_{1/2}  -  4s$&  0.6482(1)  &   $  3p_{1/2}  -  4s$&    5.997(1)  \\
$ 2p_{1/2}  -  5s$&  0.3485(1)  &   $  3p_{1/2}  -  5s$&    1.5216(5) \\
$ 2p_{1/2}  -  6s$&  0.2311     &   $  3p_{1/2}  -  6s$&    0.8023(2) \\
$ 2p_{1/2}  -  7s$&  0.1695     &   $  3p_{1/2}  -  7s$&    0.5284(2)
\\    [0.2pc]
$ 2p_{1/2} -   3d_{3/2}$&  5.0665(10)   &   $  3p_{1/2} -   3d_{3/2}$&    11.701(2) \\
$ 2p_{1/2} -   4d_{3/2}$&  1.9291(4)  &   $  3p_{1/2} -   4d_{3/2}$&    7.765(3)  \\
$ 2p_{1/2} -   5d_{3/2}$&  1.1214(4)  &   $  3p_{1/2} -   5d_{3/2}$&    3.235(1)  \\
$ 2p_{1/2} -   6d_{3/2}$&  0.7682(2)  &   $  3p_{1/2} -   6d_{3/2}$&    1.9394(8) \\
$ 2p_{1/2} -   7d_{3/2}$&  0.5736(1)  &   $  3p_{1/2} -   7d_{3/2}$&    1.3511(5) \\  [0.2pc]
$ 2p_{3/2} -    3s$&  3.4403(7)  &   $             $&              \\
$ 2p_{3/2} -    4s$&  0.9167(2)  &   $  3p_{3/2}  -  4s$&     8.481(3) \\
$ 2p_{3/2} -    5s$&  0.4929(1)  &   $  3p_{3/2}  -  5s$&     2.1518(6)\\
$ 2p_{3/2} -    6s$&  0.3268(1)  &   $  3p_{3/2}  -  6s$&     1.1347(3)\\
$ 2p_{3/2} -    7s$&  0.2397     &   $  3p_{3/2}  -  7s$&     0.7472(2)\\        [0.2pc]
$ 2p_{3/2}-     3d_{3/2} $&  2.2658(5)  &   $  3p_{3/2}  -   3d_{3/2} $& 5.233(1)     \\
$ 2p_{3/2}-     4d_{3/2} $&  0.8627(2)  &   $  3p_{3/2}  -   4d_{3/2} $& 3.473(1)     \\
$ 2p_{3/2}-     5d_{3/2} $&  0.5015(2)  &   $  3p_{3/2}  -   5d_{3/2} $& 1.4469(6)    \\
$ 2p_{3/2}-     6d_{3/2} $&  0.3435(1)  &   $  3p_{3/2}  -   6d_{3/2} $& 0.8673(3)    \\
$ 2p_{3/2}-     7d_{3/2} $&  0.2565(1)  &   $  3p_{3/2}  -   7d_{3/2} $& 0.6042(2)    \\  [0.2pc]
$ 2p_{3/2} -    3d_{5/2} $&  6.7975(14) &   $  3p_{3/2} -    3d_{5/2} $& 15.699(3)    \\
$ 2p_{3/2} -    4d_{5/2} $&  2.5882(5)  &   $  3p_{3/2} -    4d_{5/2} $& 10.418(4)    \\
$ 2p_{3/2} -    5d_{5/2} $&  1.5045(6)  &   $  3p_{3/2} -    5d_{5/2} $& 4.341(2)     \\
$ 2p_{3/2} -    6d_{5/2} $&  1.0306(2)  &   $  3p_{3/2} -    6d_{5/2} $& 2.602(1)     \\
$ 2p_{3/2} -    7d_{5/2} $&  0.7696(2)  &   $  3p_{3/2} -    7d_{5/2} $& 1.8126(7)    \\[0.2pc]
 $   3d_{3/2} -  4p_{1/2} $&    1.960(1)  & $   3d_{3/2} -  4p_{3/2} $&    0.8764(7)    \\
 $   3d_{3/2} -  5p_{1/2} $&    0.7029(4) & $   3d_{3/2} -  5p_{3/2} $&    0.3143(2)    \\
 $   3d_{3/2} -  6p_{1/2} $&    0.3984(8) & $   3d_{3/2} -  6p_{3/2} $&    0.1782(4)    \\
 $   3d_{3/2} -  7p_{1/2} $&    0.2692(4) & $   3d_{3/2} -  7p_{3/2} $&    0.1204(2)    \\    [0.2pc]
 $   3d_{5/2} - 4p_{3/2}  $&   2.629(2)   &  $         3d_{3/2} - 4f_{5/2}$&     15.82(3)          \\
 $   3d_{5/2} - 5p_{3/2}  $&   0.9430(7)  &  $         3d_{3/2} - 5f_{5/2}$&     5.141(5)          \\
 $   3d_{5/2} - 6p_{3/2}  $&   0.5345(11) &  $        3d_{5/2} - 4f_{5/2} $&    4.227(8)        \\
 $   3d_{5/2} - 7p_{3/2}  $&   0.3612(6)  &  $        3d_{5/2} - 5f_{5/2} $&    1.374(1)   \\
 $    3d_{5/2} - 4f_{7/2} $&   18.90(3)&    $    3d_{5/2} - 5f_{7/2}     $&   6.145(6)
 \end{tabular}
\end{ruledtabular}
\end{table}

 We use the linearized version of the coupled cluster approach
  (also refereed to as the all-order method), which sums infinite sets
of many-body perturbation theory terms, for $k$ with principal quantum number $n\leq 26$. The  $2s-np$, $2p-nl$, $3s-nl$, $3p-nl$, and $3d-nl$
transitions with $n\leq 26$ are calculated using this approach.
 Detailed description of the all-order
method is given in Refs.~\cite{SafJoh08,SafSaf11}.

We use experimental values of energy levels up to $n=12$ taken from
\cite{SanSimGil11,RadEngBra95,RalKraRea11} and theoretical all-order
energy levels for $n=13 - 26$. The remaining contributions with
$n>26$ are calculated in the Dirac-Fock (DF) approximation. We use a
complete set of DF wave functions on a nonlinear grid generated using
B-splines constrained to a spherical cavity.  A cavity radius of
220~$a_0$ is chosen to accommodate all valence orbitals with $n<13$
so we can use experimental energies for these states. The basis set
consists of 70 splines of order 11 for each value of the relativistic
angular quantum number $\kappa$.

 The evaluation of the uncertainty of the matrix elements in this approach was
described in detail in \cite{SafSaf11}. Four all-order calculations were
 carried out. Two of these were \textit{ab initio} all-order calculations with
and without the inclusion of the partial triple excitations. Two
other calculations included semiempirical estimate of high-order
correlation corrections starting from both \textit{ab initio} runs.
The spread of these
 four values for each transition defines the  estimated uncertainty in the final
results. Since high-order corrections are small for some Li
transitions, in particular for the excited states, the uncertainty
estimate for the lower transition  was used for the other
transitions. For example, the  $2p-3s$ uncertainty was used for the
other  $2p-ns$ transitions with $n>3$. This procedure insures that
all uncertainty estimates are larger than the expected numerical
accuracy of the calculations. The uncertainties for the small
$2s-np_j$ and $3s-n_1p_j$ matrix elements with $n>2$ and $n_1>3$ were
estimated as 10\% of the total correlation correction. For these
transitions, the procedure used to estimate high-order corrections
does not estimate all dominant contributions. Placing the uncertainty
estimate at 10\% of the correlation correction for these transitions
ensures that all missing correlation effects do not exceed this rough
uncertainty estimate. The matrix element calculations have been
carried out for a fixed nucleus using a Fermi distribution for a
nuclear charge. These matrix elements are used in all polarizability
calculations in the present work.  We have estimated that the $^6$Li
- $^7$Li isotope shift correction to the matrix elements is well
below our estimated uncertainties.

The absolute values of the reduced electric-dipole Li matrix elements
 used in our subsequent calculations and their uncertainties are listed in a.u.
in Table~\ref{tab1}. We note that we list only the  most important subset
of the matrix elements. We have calculated a total of 474 matrix elements
for this work.

\begin{table}[ht]
\caption[]{Comparison of the present values of static scalar
$\alpha_0$ and tensor $\alpha_2$ $^7$Li polarizabilities with other
calculations and experiment. The final entry is the difference of the
$2p_{1/2}$ and $2s$ polarizabilities for which there is a precise
experimental determination using dc Stark shift measurement. Units:
$a^3_0$.} \label{tab2}
\begin{ruledtabular}
\begin{tabular}{llll}
   \multicolumn{1}{c}{State} & \multicolumn{1}{c}{Ref.} &
  \multicolumn{1}{c}{$\alpha_0$} & \multicolumn{1}{c}{$\alpha_2$} \\
\hline
  $2s$       & Present &  164.16(5)&\\
            & Hylleraas \cite{TanYanShi10}&  164.11(3)&\\
            & Hylleraas \cite{PucKedPac11,PucKcePac12}&  164.1125(5)&\\
            & CCSD(T) \cite{WanSahTim10}& 164.23 &\\
            & Expt. \cite{MifJacBuc06} & 164.2(11)&\\[0.3pc]
  $2p_{1/2}$ & Present&  126.97(5)&\\
            & CCSD(T) \cite{WanSahTim10}& 127.15 &\\
  $2p_{3/2}$ &Present&  126.98(5) &   1.610(26)\\
           & CCSD(T) \cite{WanSahTim10}& 127.09 & 1.597\\
  $2p$       &Hylleraas \cite{TanYanShi10}&126.970(4)& 1.612(4)\\[0.3pc]
  $3s$       &Present&   4130(1)  &\\
             & Hylleraas  \cite{TanZhaYan10}& 4131.3\\
             & CICP \cite{TanZhaYan10} & 4135 &\\
        & CCSD(T) \cite{WanSahTim10}& 4116 & \\ [0.3pc]
  $3p_{1/2}$ &Present&  28255(11) &\\
          & CCSD(T) \cite{WanSahTim10}& 28170 & \\
  $3p_{3/2}$ &Present& 28261(10) & -2170(2)\\
          & CCSD(T) \cite{WanSahTim10}& 28170 & -2160 \\
    $3p$     & Hylleraas \cite{TanZhaYan10}& 28250 & -2168.3\\
        & CICP \cite{TanZhaYan10} & 28450 &-2188\\[0.3pc]
  $3d_{3/2}$ &Present& -14925(8)  & 11405(6)\\
          & CCSD(T) \cite{WanSahTim10}& -14820 & 11460\\
  $3d_{5/2}$ &Present& -14928(9)  & 16297(7)\\
          & CCSD(T) \cite{WanSahTim10}& -14930& 16290\\
  $3d$       &Hylleraas \cite{TanYanShi09}& -14928.2 & 16297.7\\
  $3d$       &Expt. \cite{AshClaWij03}&-15130(40) & 16430(60)\\[0.3pc]
    $2p_{1/2}-2s$ &Present&    -37.19(7)\\
           & CCSD(T) \cite{WanSahTim10}& -37.09 & \\
                &Expt.\cite{HunKraBer91} & -37.15(2)
 \end{tabular}
\end{ruledtabular}
\end{table}

\section{Li polarizabilities}
\label{sec2}
 We start with a brief overview of the recent calculations of Li polarizabilities.
 The
highest accuracy attained in \textit{ab initio} atomic structure
calculations is achieved by the use of basis functions which
explicitly include the interelectronic coordinate. Difficulties with
performing the accompanying multi-center integrals have effectively
precluded the use of such basis functions for systems with more than
three electrons. Correlated basis calculations are possible for
lithium since it only has three electrons. Consequently it has been
possible to calculate Li polarizabilities to very high precision
\cite{TanYanShi09,TanYanShi10,TanZhaYan10,PucKedPac11}. The most
accurate calculation of  polarizability of  lithium ground state
 $\alpha_0(2s)$ = 164.1125(5)~a.u. was carried out in
\cite{PucKedPac11}. This calculation included  relativistic and
quantum electrodynamics corrections. The small uncertainty comes from
the approximate treatment of quantum electrodynamics corrections.
This theoretical result can be considered as a benchmark for more
general atomic structure methods and may serve as a reference value
for the relative measurement of polarizabilities of the other
alkali-metal atoms \cite{PucKedPac11,HolRevLon10}. The uncertainty in
the experimental value of the polarizability 164.2(11) a.u.
\cite{MifJacBuc06} is substantially higher. The most stringent
experimental test of Li polarizability calculations is presently the
Stark shift measurement of the $2s$-$2p_{1/2}$ transition  by Hunter
{\em et al.} \cite{HunKraBer91}, which gave a polarizability
difference of $-$37.15(2) a.u.   Our coupled-cluster result
$-$37.19(7)~a.u. is in excellent agreement with the experimental
polarizability difference.

The electric dipole polarizabilities and hyperpolarizabilities for the lithium isotopes
$^6$Li and $^7$Li in the $2s$,  $2p$, and $3d$ excited states
were calculated nonrelativistically using the variational method with a Hylleraas basis \cite{TanYanShi09}. The calculated $3d$ polarizabilities were
found to be in significant disagreement with 2003 measurements \cite{AshClaWij03}. In order resolve this discrepancy, we have calculated $3d_j$
static scalar and tensor polarizabilities. Our results are in excellent agreement with the Hylleraas calculations of \cite{TanYanShi09} indicating a
problem with measurements reported in \cite{AshClaWij03}.
 The dc and ac dipole polarizabilities for Li atoms  in the $2s$ and $2p$ states were calculated using the same variational method
with a Hylleraas basis in \cite{TanYanShi10}. Corrections due to relativistic effects were also estimated to provide recommended values.

Since the Hylleraas basis set calculations can not be carried out for larger system at
 the present time, Li provides an excellent benchmark test of
our coupled-cluster approach, and our procedures to estimate the
uncertainties. Our approach is intrinsically relativistic unlike
nonrelativistic Hylleraas calculations that requires subsequent
estimate of the relativistic corrections. We note that the
coupled-cluster method used in this work is not expected to provide
accuracy that is competitive with what is ultimately possible to
achieve with Hylleraas basis for dc polarizability results. The
accuracy of the coupled-cluster method was recently discussed in
\cite{DerPorBel08}. However, the coupled-cluster approach is
applicable to more complicated systems, so comparison with Hylleraas
calculations in Li provides stringiest tests of the methodology as
well as of numerical accuracy. Moreover, comparison with Hylleraas
benchmarks provides additional confidence in our estimates of the
uncertainties in the values of magic wavelengths. Comparison of our
results with most recent correlated basis set calculations
\cite{TanYanShi09,TanYanShi10,TanZhaYan10,PucKedPac11,PucKcePac12} is
given in Table~\ref{tab2}. All data are given for $^7$Li since the
differences between $^6$Li and $^7$Li values are smaller than  our
uncertainties. Our results are in excellent agreement with all
Hylleraas values.

\begin{figure}
  \includegraphics[width=2.7in]{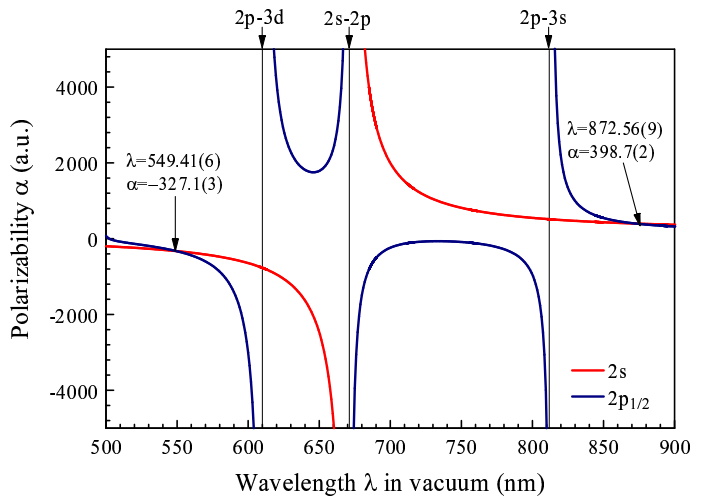}
  \caption{(Color online) The frequency-dependent polarizabilities of the Li $2s$ and $2p_{1/2}$ states.
  The magic wavelengths are marked with arrows.}
  \label{fig1}
\end{figure}
\begin{figure}
  \includegraphics[width=2.7in]{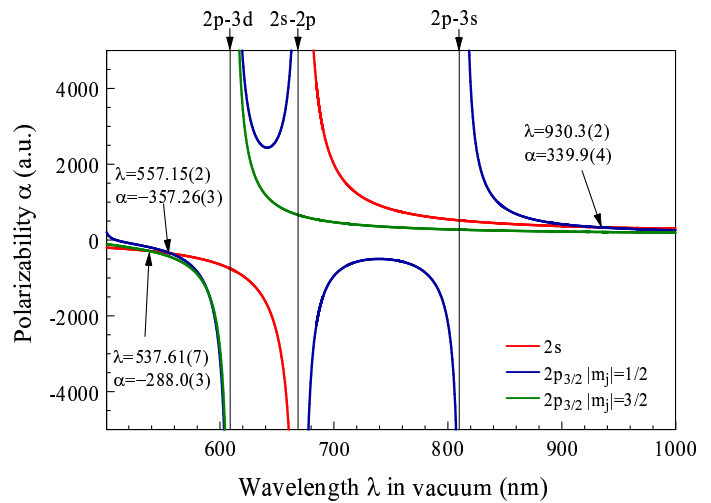}
  \caption{(Color online) The frequency-dependent polarizabilities of the Li $2s$ and $2p_{3/2}$ states.
  The magic wavelengths are marked with arrows.}
  \label{fig2}
\end{figure}

Li electric-dipole and quadrupole polarizabilities, and van der Waals
 coefficients were calculated using relativistic coupled-cluster method with
 single, double, and partial triple excitations, CCSD(T), in
Ref.~\cite{WanSahTim08}. However, a mismatch of phases between the numerical and analytical orbitals caused significant errors in the reported
values. The later revision of this work \cite{WanSahTim10}, where this problem was corrected, still yielded results that are in substantially poorer
agreement with Hylleraas data than the values in the present work.
 The importance of using very accurate basis sets
for coupled-cluster calculations of polarizabilities was discussed in
a number of publications including review \cite{MitSafCla10}. In our
work, very large basis set with 70 orbitals for each partial wave was
used resulting in a very high numerical accuracy.
 Table~\ref{tab2} includes the comparison with the revised CCSD(T)  calculation of
\cite{WanSahTim10}, and configuration interaction calculations with
core polarization (CICP) \cite{TanZhaYan10}.

\begin{figure}
  \includegraphics[width=2.7in]{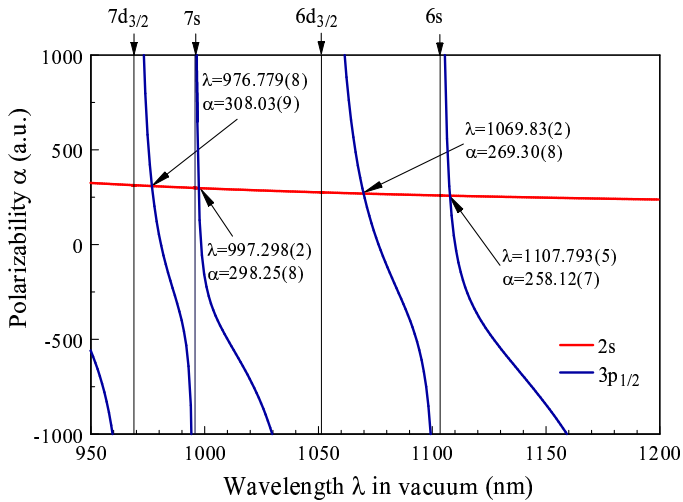}
  \caption{(Color online) The frequency-dependent polarizabilities of the Li $2s$ and $3p_{1/2}$ states.
  The magic wavelengths are marked with arrows.}
  \label{fig3}
\end{figure}
\begin{figure}
  \includegraphics[width=2.7in]{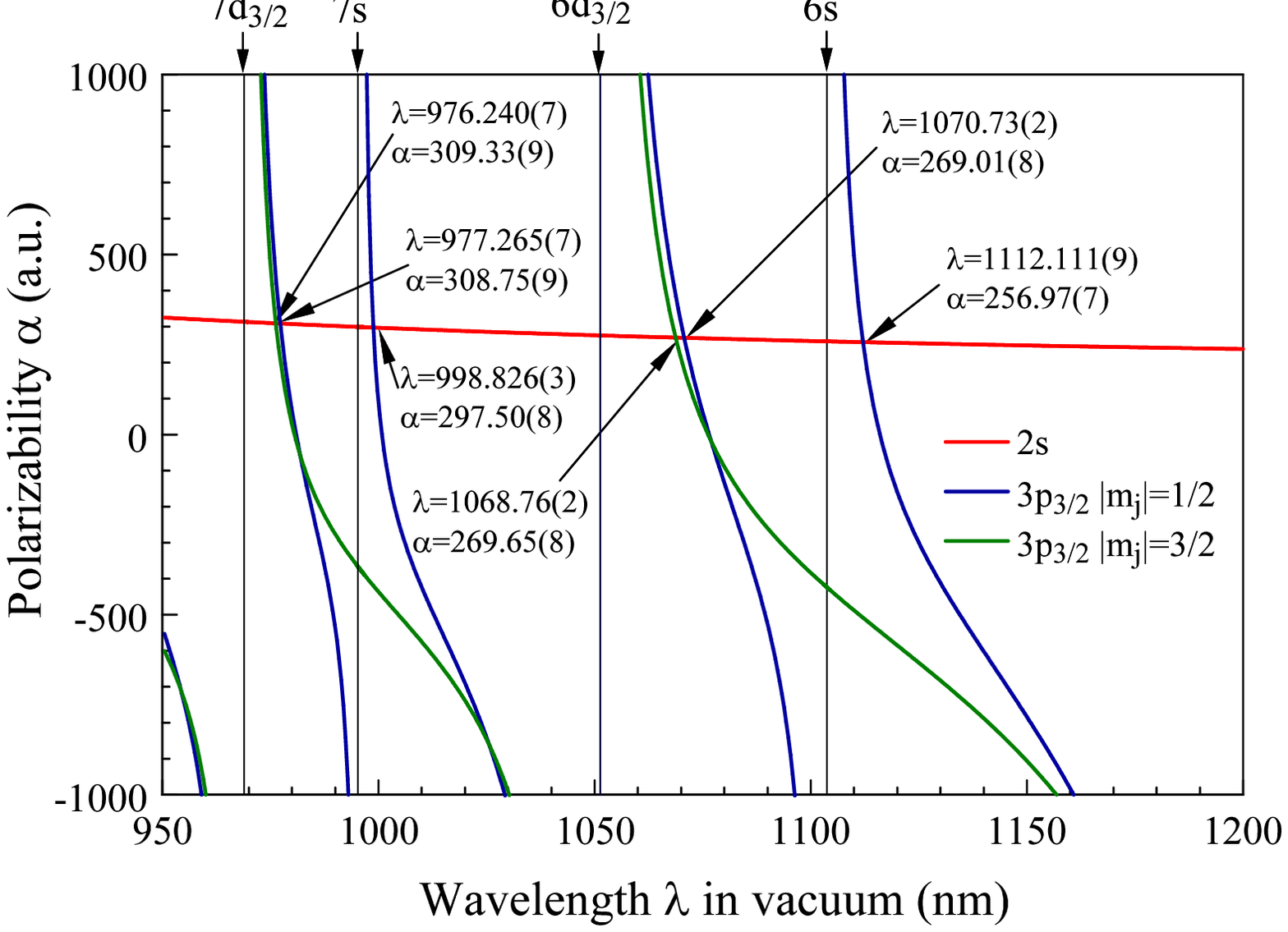}
  \caption{(Color online) The frequency-dependent polarizabilities of the Li $2s$ and $3p_{3/2}$ states.
  The magic wavelengths are marked with arrows.}
  \label{fig4}
\end{figure}

\begin{table}[ht]
\caption[]{Magic wavelengths for the $2s-np_j$ transitions in $^6$Li
and $^7$Li. The $400-950$~nm and $950-1300$~wavelength ranges were
considered for the $2s-2p_j$ and $2s-3p_j$ transitions, respectively.
The corresponding polarizabilities are given in a.u. } \label{tab3}
\begin{ruledtabular}
\begin{tabular}{lrrr}
   \multicolumn{1}{c}{Transition} & \multicolumn{1}{c}{$^6$Li $\lambda_\textrm{magic}$} &    \multicolumn{1}{c}{$^7$Li $\lambda_\textrm{magic}$} &
  \multicolumn{1}{c}{$\alpha$} \\
\hline
$2s-2p_{1/2}$&     401.247(7)  &401.245(7) &    -87.07(4)  \\
             &     425.801(3)  &425.798(3) &   -106.32(4)  \\
             &     434.84(1)   &434.84(1)  &   -114.46(5)  \\
             &     494.738(3)  &494.735(3) &   -190.40(6)  \\
             &     549.41(6)   &549.42(6)  &   -327.1(3)   \\
             &     872.56(9)   &872.57(9)  &    398.7(2)   \\  [0.2pc]
$2s-2p_{3/2}$ &   402.149(2)   &402.146(2) &     -87.71(3)   \\
$|m_j|=1/2$   &   425.142(2)   &425.139(2) &    -105.75(3)   \\
              &   436.827(4)   &436.825(4) &    -116.33(3)   \\
              &   493.351(1)   &493.348(1) &    -188.04(5)   \\
              &   557.15(2)    &557.16(2)  &   -357.26(7)    \\
              &   930.3(2)     &930.3(2)   &     339.9(2)    \\    [0.2pc]
  $2s-2p_{3/2}$ &   431.46(2)  &431.45(2)  &   -111.33(5)  \\
  $|m_j|=3/2$   &   537.61(7)  &537.61(7)  &   -288.0(3)   \\    [0.2pc]
$2s-3p_{1/2}$ &      976.779(8)&976.777(8)  &     309.03(9)  \\
              &      997.298(2 &997.295(2)  &     298.25(8)  \\
              &     1069.83(2) &1069.82(2)  &    269.30(8)   \\
              &     1107.793(5)&1107.783(5) &     258.12(7)  \\
              &     1287.25(6) &1287.24(6)  &    224.66(7)   \\    [0.2pc]
$2s-3p_{3/2}$ &    977.265(7)  &977.264(7)  &   308.75(9) \\
$|m_j|=1/2$   &    998.826(3)  &998.823(3)  &   297.50(8) \\
              &    1070.73(2)  &1070.72(2)  &   269.01(8) \\
              &   1112.111(9)  &1112.102(8) &   256.97(7) \\
              &   1288.15(4)   &1288.15(4)  &  224.55(7)  \\    [0.2pc]
 $2s-3p_{3/2}$ &   976.240(7)  &976.239(7)  &   309.33(9)  \\
 $|m_j|=3/2$   &  1068.76(2)   &1068.74(2)  &  269.65(8)   \\
               &  1285.88(6)   &1285.87(6)  &  224.84(7)
  \end{tabular}
\end{ruledtabular}
\end{table}

\section{Magic wavelengths}
\label{sec3}

We define the magic wavelength $\lambda_{\rm{magic}}$ as the wavelength
 for which the ac polarizabilities of  two states involved in the atomic transition
  are the same, leading to
vanishing ac Stark shift of that transition. For the $ns-np$
transitions, a magic wavelength is represented by the point at which
two curves, $\alpha_{ns}(\lambda)$ and $\alpha_{np}(\lambda)$,
intersect as a function of the wavelength $\lambda$.

The total polarizability of the $np_{3/2}$ state depends
 upon its $m_j$ quantum number. Therefore,
  the magic wavelengths need to be determined separately for the cases with $m_j=\pm 1/2$ and
$m_j=\pm 3/2$  for the $ns-np_{3/2}$ transitions, owing to the
presence of the tensor contribution to the total polarizability of
the $np_{3/2}$ state. The total polarizability for the $np_{3/2}$
states is given by  $\alpha=\alpha_0-\alpha_2$ for $m_j=\pm 1/2$ and
$\alpha=\alpha_0+\alpha_2$  for the $m_j=\pm 3/2$ case.
 The uncertainties in the values of magic wavelengths are found as the maximum differences between the central value and the crossings of the
$\alpha_{ns} \pm \delta \alpha_{ns}$ and $\alpha_{np} \pm \delta \alpha_{np}$
 curves, where the  $\delta \alpha$
are the uncertainties in the corresponding $ns$ and $np$ polarizability values.
 All calculation are carried out for linear polarization.
\begin{table} [h]
\caption{\label{tab5}  Tune-out wavelengths $\lambda_{\rm{zero}}$ for
$^6$Li and $^7$Li. The resonant wavelengths $\lambda_{\rm{res}}$ for
relevant transitions are also listed. The wavelengths (in vacuum) are
given in nm.}
\begin{ruledtabular}
\begin{tabular}{clll}
\multicolumn{1}{l}{Atom}& \multicolumn{1}{l}{Resonance}&
\multicolumn{1}{l}{$\lambda_{\rm{res}}$}&
\multicolumn{1}{l}{$\lambda_{\rm{zero}}$}\\
\hline \\[-0.2pc] $^6$Li&$2s-2p_{1/2}$     & 670.992478
 &            \\
&                  &       & 670.987445(1)\\
&$2s-2p_{3/2}$     & 670.977380&  \\[0.2pc]
&                  &       & 324.19(2)\\
&$2s-3p_{1/2}$ & 323.3622 & \\
&$2s-3p_{3/2}$ &  323.361168 & \\
&                  &       & 274.920(10) \\
&$2s-4p_{1/2}$ & 274.2035
& \\[0.2pc]

\hline \\[-0.2pc]  $^7$Li&$2s-2p_{1/2}$     & 670.976658
 &            \\
&                  &       & 670.971625(2)\\
&$2s-2p_{3/2}$     & 670.961561&\\
&                  &       & 324.19(2)\\
&$2s-3p_{1/2}$ & 323.3572 & \\
&$2s-3p_{3/2}$ & 323.3562 & \\
&                  &       & 274.916(10) \\
&$2s-4p_{1/2}$ & 274.1998&
\end{tabular}
\end{ruledtabular}
\end{table}
The frequency-dependent polarizabilities of the Li ground and
$2p_{1/2}$ states for $\lambda=500 - 900$~nm  are plotted in
Fig.~\ref{fig1}. The positions of the resonances are indicated by
vertical lines with small arrows on top of the graph. There are only
3 resonances ($2p-3d$, $2s-2p$, and $2p-3s$) in this wavelength
region, resulting in only two magic wavelengths above 500~nm for the
$2s-2p_{1/2}$ transition. These occur at 549.41(6)~nm and
872.56(9)~nm.
 The magic wavelengths are marked with arrows on all graphs.
  The respective values of the polarizabilities are also  given for all magic
  wavelengths
  illustrated in this work. There are no other magic wavelengths for $\lambda > 900$~nm
   since there are no more resonance contributions
   to both $2s$ and $2p_{1/2}$
polarizabilities in this region. Both polarizabilities slowly
decrease for $\lambda>900$~nm to approach their static value in the
$\omega \longrightarrow 0$ limit.

The frequency-dependent polarizabilities of the Li ground and
$2p_{3/2}$ states for $\lambda=500 - 900$~nm are plotted in
Fig.~\ref{fig2}. The same designations are used in all figures. The
magic wavelengths for the $2s-2p_{3/2} ~|m_j|=1/2$ transition are
similar to the ones for the $2s-2p_{1/2}$ transition. In the
$|m_j|=3/2$ case, there is no magic wavelength above 550~nm since
 there is no contribution from the $2s-2p$ resonance.

The magic wavelengths for the $2s-3p_{1/2}$ and $2s-3p_{3/2}$ transitions for $\lambda=950 - 1200$~nm are illustrated in Fig.~\ref{fig3} and
Fig.~\ref{fig4}. All resonances indicated on top of the figures refer to the $3p-nl$ transitions, i.e. arrow labelled $7s$ indicates the position of
the $3p-7s$ resonance. A number of the magic wavelengths are available for the $2s-3p$ uv transitions owing to a large number of resonance
contributions to the $3p$ polarizabilities in this region. The $3s-nl$ resonances with $n>7$ will yield other magic wavelengths to the left of the
plotted region, while the $3s-nl$ resonances with $n<6$ will yield other magic wavelengths to the right of the plotted region.

We have also calculated the ac polarizability of the $3s$ state for
$\lambda=950 - 1200$~nm. We find that $3s$ polarizability is negative
and large ($\approx$-1000~a.u.) in this entire wavelength region with
the exception of the very narrow wavelength interval between 1079.48
- 1079.50~nm (due to $3s-4p_j$ resonances near 1079.49~nm).

Table~\ref{tab3} presents the magic wavelengths for the $2s-2p_j$
transitions in the $400-950$~nm region, and
 the magic wavelengths for the $2s-3p_j$ transitions in the
$950-1300$~nm region. We have carried out separate calculations for
$^6$Li and $^7$Li by using the experimental energies for each isotope
from \cite{SanSimGil11,RadEngBra95}. The same theoretical matrix
elements were used for both isotopes, since the isotopic dependence
of the  matrix elements is much less than out uncertainty.
 We find
that the differences between $^6$Li and $^7$Li magic wavelengths is
very small and is smaller than our estimated uncertainties for almost
all of the cases. We list both sets of data for the illustration of
the IS differences. The polarizabilities at the magic wavelengths for
$^6$Li and $^7$Li are the same within the listed uncertainties, so
only one set of polarizability data is listed.

While our calculations are not sensitive enough to significantly
differentiate between magic wavelengths for $^6$Li and $^7$Li, our
values of the first tune-out wavelength are significantly different
for the two isotopes ~\cite{AroSafCla11}. Therefore, we consider Li
tune-out wavelengths in more detail in the next section.

\section{Li tune-out wavelengths}

A tune-out  wavelength $\lambda_{\rm{zero}}$ for a given state is one
for which the
 ac polarizability of that state vanishes.  In practice, we calculate $\alpha_0(\omega)$
 for a range of frequencies in the vicinity of  relevant resonances and
look for changes in sign of the polarizability within a given range.
In the vicinity of the sign change, we vary the frequency until the
residual polarizability is smaller than  our uncertainty.

\begin{figure}
  \includegraphics[width=2.6in]{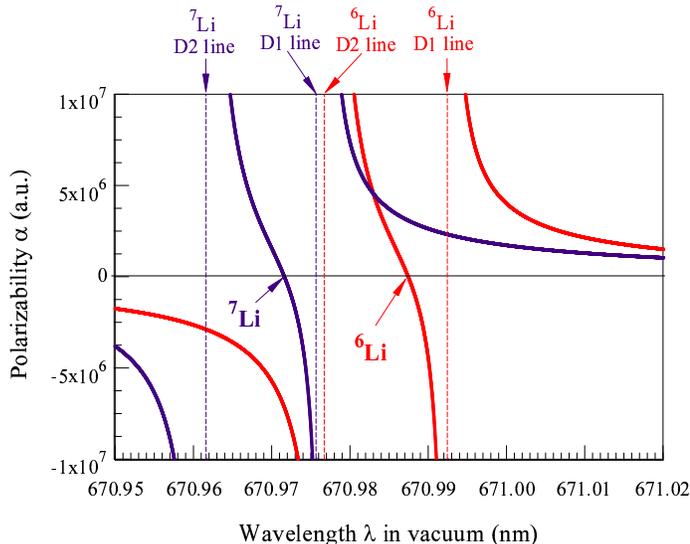}
  \caption{(Color online) The frequency-dependent polarizabilities of the ground state
  of $^6$Li and $^7$Li.
  The tune-out wavelengths are marked with arrows.}
  \label{fig5}
\end{figure}

In previous work ~\cite{AroSafCla11}, we presented the tune-out
wavelengths for the ground state of Li. Here, we
 have calculated the matrix elements to somewhat higher
precision and have carried out more extensive study of their
uncertainties. Therefore, we reevaluate the tune-out wavelengths for
Li in the present work.  We have carried out separate calculations
for $^6$Li and $^7$Li by using the experimental energies  for each
isotope from \cite{SanSimGil11,RadEngBra95} but the same matrix
elements. Only the values of the first tune-out wavelengths are
significantly affected by the IS. We illustrate this case separately
in Fig.~\ref{fig5}. The isotope shift of the $2s-2p$ line is slightly
larger than its fine structure. Therefore, the tune-out wavelength
for one isotope corresponds to a very deep trap for the other. We
note that the first tune-out wavelength is very close to the D1, D2
resonances which would be a source of strong light scattering in an
experiment. Applications of the tune-out wavelengths to sympathetic
cooling and precision measurement were discussed in
\cite{AroSafCla11}.

\section{Conclusion}
We have calculated the ground $ns$ state and  $np$ state ac
polarizabilities in Li using the relativistic linearized
coupled-cluster method and evaluated the uncertainties of these
values. The static polarizability values were found to be in
excellent agreement with experiment
 and high-precision theoretical
calculations with correlated basis functions. We have used our
calculations
 to identify the magic wavelengths for the $2s-2p$ and $2s-3p$
 transitions relevant to the use of ultraviolet resonance lines for
 laser cooling of ultracold gases with high phase-space densities.

\section*{Acknowledgement}
This research was performed under the sponsorship of the US Department of Commerce, National Institute of Standards and Technology, and was supported
by the National Science Foundation under Physics Frontiers Center Grant PHY-0822671.
\newpage

\end{document}